\documentclass[12pt]{article}
\usepackage{graphicx}
\begin{document}
\begin{center}
{\Large\bf Determining the neutron star equation
of state using the narrow-band gravitational wave detector
Schenberg}

\vspace{1cm}

{\it Guilherme F Marranghello and
 Jos\'{e} C N de Araujo}

\vspace{1cm}

{\bf Instituto Nacional de Pesquisas Espaciais -
Divis\~ao de Astrof\'\i sica \\ Av. dos Astronautas 1758, S\~ao
Jos\'e dos Campos, 12227-010 SP, Brazil}
\end{center}

\begin{abstract}
We briefly review the properties of quasi-normal modes of neutron
stars and black holes. We analyze the consequences of a possible
detection of such modes via the gravitational waves associated with
them, specially addressing our study to the Brazilian spherical
antenna, on which a possible detection would occur at 3.0-3.4 kHz. A
question related to any putative gravitational wave detection
concerns the source that produces it. We argue that, since the
characteristic damping times for the gravitational waves of neutron
stars and black holes are different, a detection can distinguish
between them; and also distinguish the neutron star's oscillating
modes. Moreover, since the source can be identified by its
characteristic damping time, we are able to extract information
about the neutron star or black hole. This information would lead,
for example, to a strong constrain in the nuclear matter equation of
state, namely, the compression modulus should be $K\approx 220 MeV$.
\end{abstract}


04.30.Db, 04.80.Nn, 95.55.Ym, 97.60.Jd, 97.60.Lf



\section{Introduction}
A new window to the Universe is about to be opened. With the
detection of gravitational waves (GWs), astrophysicists expect to
have the answer to many questions, as well as new ones. There are
many detectors all around the world and, soon, even above us. With
interferometric or resonant mass detectors, at low and high
frequencies, we shall be able to see waves coming from coalescing
binary systems, from a cosmological background, from catastrophic
events etc.

Among the most promising sources of GWs, neutron stars (NSs) and
black holes (BHs) are the best candidates for a detection. These
objects emit waves in a very wide spectrum of frequencies determined
by their quasi-normal modes of oscillation; and many works have been
recently done to determine their characteristics (see
Ref.\cite{kokkotas} for a review).

We take a special attention to a small region of this spectrum,
localized at 2.8-3.4 kHz, which is the region of operation of the
resonant spherical antennas Mario Schenberg (Brazilian, 3.0-3.4 kHz)
\cite{aguiar} and Mini-GRAIL (Dutch, 2.8-3.0 kHz) \cite{waard}. Both
antennas are composed of  $\sim 1\, {\rm ton}$ and $\sim 70\, {\rm
cm}$ diameter CuAl6$\%$ spheres and will be operating in the
frequency range mentioned above.

It is worth stressing that the identification of the direction of a
source, making use of only one detector, is possible only if it is a
spherical antenna \cite{aguiar}. Several interferometer and bar
detectors would be necessary to do the same.

One could ask how well a spherical antenna can determine the
direction of a source. The Schenberg antenna operating at $T\sim 20
mK$, for example, may have an angular resolution of  $\theta\approx
2^o$ \cite{costa2006}. With the help of any electromagnetic data it
is possible to locate and identify a given source.

When we take such a small window (2.8-3.4 kHz) to such a giant
garden named the Universe, we are sure that we are losing a great
amount of information. However, if we are able to focus on good
accuracy to this small region, we can also expect to see all the
magnificence of it. This is the case we are proposing in this work.

We analyze the case of a possible future detection made by these
antennas in order to obtain information on its GW sources. In the
case of BHs, a direct detection of these mysterious objects is
already of great relevance, but we are specially concerned to the
information we can extract from a NS detection. If values of mass
and radius can be extracted with some accuracy, we could get
additional information about the nuclear matter equation of
state\cite{kok}. The still obscure characteristics of nuclear
matter, e.g., the values of compression modulus or nucleon effective
mass could be clarified. We could also get information on the
strange quark star candidates, strange quark matter equation of
state and phase transition\cite{marranghello}.

A relevant and actual issue in astrophysics refers to the existence
and detection of strange quark stars. Some objects have been
recently detected and issued as possible candidates to be stars
formed by deconfined quark matter. This is the case of  EXO
0748-676, 4U 1700+24, 1E1207-5209 and RX J1856.5-3574
\cite{bagchi,drake}. We do not intend, in this work, to discuss the
real constitution of these stars, we just intend to show how a GW
detection could identify a star as a strange star.

This paper is organized as follows. In section 2 we briefly review
the BH and NS pulsating modes, in section 3 we discuss the
detectability of NS modes by the Brazilian antenna, in section 4 we
consider the constraints a putative detection by this very antenna
would impose on a class of equation of states, finally in section 5
we present the final remarks.

\section{Black Holes and Neutron Stars}

NSs and BHs are certainly the most important sources of GWs, which
can be generated by a host of pulsating modes of these stars.

While BHs oscillate due to its spacetime structure perturbation, NSs
may also oscillate due to its fluid perturbation. This difference
implies that a NS may oscillate in many different modes and have a
richer spectrum of oscillations.

An excited BH may emit a great amount of GWs during its ringdown
phase. The ringdown waveform may be completely described by the BH's
mass and angular momentum.

Although NSs possess many pulsation modes only a few of them,
however, is of relevance for GW detection. Concerning the GW point
of view the most important modes for the NSs are the fundamental (f)
mode of fluid oscillation, the first few pressure (p) modes and the
first GW (w) mode \cite{kokkot1992}.

It is worth noting that different types of initial perturbations are
more convenient to excite each of these modes \cite{ruoff}. The
problem of initial value in numerical relativity is one of the most
challenging problems to be solved in the present. However, there is
a currently belief that the fundamental mode (f-mode) may carry the
greatest part of the released energy, because the fluid parameters
undergo the largest changes, while the pressure (p-mode) and
curvature (w-mode) would carry smaller fraction of the energy.

It is worth mentioning that the r-mode can also be, under certain
circumstances, a very important source of GWs for the NSs
\cite{anders1999}.

An important question is how pulsating modes are excited in the NSs,
which are of our concern here. There are many scenarios that could
lead to significant asymmetries. Supernova explosions are expected
to form wildly pulsating NSs that emit GWs. A pessimistic estimate
for the energy radiated as GWs indicates a total release equivalent
to $< 10^{-6}M_{\odot}$. An optimistic estimate, when the NS is
formed, for example, from strongly non-spherical collapse, suggests
a release equivalent to $10^{-2} - 10^{-1} M_{\odot}$. (see, e.g.,
Refs. \cite{kokkotas,kok,marranghello,pons,horvath,lin}.

Another possible excitation mechanism for NS pulsation is a
starquake, which can be associated with a pulsar glitch. The energy
released in this process may be of the order of the maximum
mechanical energy stored in the crust of the NSs, which is estimated
to be of $10^{-9} - 10^{-7}M_{\odot}$ \cite{blaes1989,mock1998}. In
a recent study \cite{araujo2005} we consider the detectability of
f-mode NSs in this scenario by the Brazilian antenna Schenberg. We
showed that several events every year could be detectable by this
antenna.

Soft gamma ray repeaters could be thought as starquakes occuring in
magnetars. This  process would be responsible for an energy release,
in the form of GWs, about to $E\sim 10^{-6}M_\odot c^2$, being
detectable if it is in our own galaxy \cite{papa}.

Orbiting or falling masses into NSs have been recently studied as
models for coalescing or acreting binaries \cite{pons}. Stellar
oscillations being excited by the tidal fields of the two stars, for
example.

Core  quakes due to phase transitions in the inner shells of the NS
could be  able to excite pulsation modes, releasing about
$E\sim10^{-2}-10^{-1}M_\odot c^2$ in the form of GWs, and could be
detected up to the Virgo cluster \cite{marranghello,lin}. Similarly,
the transformation of a NS into a strange star is likely to induce
pulsations.

Also, in ref.\cite{ma}, the rate of phase transition inside NSs has
been estimated to be about $10^{-6}$ events per year and per galaxy.
If in fact a detection could be possible up to the Virgo cluster,
this would represent a possible source of GWs.

Excitation of f-mode in LMXBs is also possible, and in this case the
GW could well be recurrently produced, since the NS is continuously
receiving matter from its companion star.  In a recent paper de
Araujo {\it etal} \cite{araujo2006} study such a case.

Since the f-mode unstable NSs in LMXBs could well be a recurrent
source of GWs, this implies that a non negligible event rate could
occur in this case. A rough estimate, based on the event rate of
superbursts in LMXBs, indicates that two events every year could be
detected if a sensitivity for burst sources of $h_{s}\sim 10^{-23}$
could be achieved, and also if the efficiency of generation of GWs
could be of $\varepsilon_{GW}\sim 10^{-11}$.

Finally, it is worth mentioning that Kokkotas {\it etal}
\cite{kokkot2001b} shows that detecting the f-mode, the EOS, the mass
and the radius of the NSs will be strongly constrained. The reader
should appreciate the reading of this paper by Kokkotas {\it etal} who
show in detail how these above mentioned astrophysical information
is obtained from the GW data.

In the following subsection we briefly review the main properties of
the NSs pulsating modes.

\subsection{f- and p-modes}

The pressure modes are the overtones of the fundamental mode. The
fundamental mode can be described by the density distribution inside
the star, while the p-mode restoration force is the pressure. The
f-mode, for a typical NS, may appear with a GW frequency of
$\omega\sim 2.8kHz$ and damping time of $\tau\sim 0.1 s$ while the
first p-mode appears with $\omega\sim 6 kHz$ and $\tau\sim 0.6 s$
\cite{kokkotas}.

The fraction of energy released by each mode of oscillation is still
an open problem. Only full 3D relativistic numerical
simulations,with appropriate initial conditions, could give us
considerable new information about this issue. Even though, it is a
common belief, as already mentioned, that the fundamental mode is
more likely to be excited than the pressure modes. It would be
required tens to hundred times more powerful events to make the
detection of a pressure mode comparable to the fundamental one
\cite{kokkotas}. Higher pressure modes shall not contribute
considerably to the GW spectrum.

At the actual stage of GW detectors it would be required an
effective amplitude about $h_{eff}\sim 10^{-21}$ to claim for a
detection. The effective amplitude for an f-mode, described by
Kokkotas \cite{kokkotas} is given by

\begin{equation}
h_{eff}\sim 2.2\times 10^{-21}\left(\frac{E}{10^{-6}M_\odot c^2}\right)^{1/2}
\left(\frac{2 kHz}{f}\right)^{1/2} \left(\frac{50 kpc}{r}\right) \, \, ,
\end{equation}
where $E$ is the energy released in the mode, $f$ is the frequency
in $kHz$ and $r$ is the distance to the source in $kpc$.

In Ref. \cite{benhar} the authors have calculated the properties of the
oscillation modes using a wide sample of equations of state. As the main
results, they have obtained empirical formulae for, among others, the
frequency of the f- and first p-mode

\begin{equation}
\nu_f=0.79(\pm0.09)+33(\pm2)\sqrt{\frac{M}{R^3}} \, \, ,\label{fr}
\end{equation}

\begin{equation}
\nu_p=\frac{1}{M}\left[-1.5(\pm0.8)+79(\pm4)\frac{M}{R}\right] \, \, .
\end{equation}
where the masses and radii are given in km, while $\nu_f$ and
$\nu_p$ are given in kHz.

Inverting the problem and considering the figures to the Schenberg
(3.0-3.4kHz) and Mini-GRAIL (2.8-3.0kHz) detectors bandwidths, one
sees the kind of information it is possible to obtain by using the
above equations.

Assuming a putative detection of a standard $1.4M_\odot$ NS, a very
small radius, namely, $R\sim 7.3 km$, would be inferred, which is
too small for a standard NS.

As another example, if a solar mass f-mode unstable NS is detected,
the radius of the star would be constrained to $R=6.58\pm 0.79 km$.
Given that this radius is too small for a NS, it would be certainly
a strange quark star.

Note that a NS of $M\sim 1.0\,( 1.4)\,M_\odot$ with $R\sim 14\, km$
would emit waves in much lower frequencies, $\nu_f\sim 1.5\,
(1.7)\,kHz$.

Also, if a detection is done and identified as a standard
$1.4M_\odot$ NS p-mode unstable, a too large radius would be
inferred, namely, $R\sim 20\, km$, which is inconsistent with any NS
model.

If the  detection points to a putative half solar mass NS p-mode
unstable, it would lead to a radius of $R=16.26\pm 4.75 km$. This
result could hardly, by itself, rule out any of the present
equations of state, specially considering that the p-mode damping is
very dependent on the stellar model and cannot be well fitted as in
the case of the f-mode (see Ref.\cite{benhar} for details). Due to
its frequency bandwidth, the Schenberg detector could not see NSs
far from this mass range.

NSs have a very rich spectrum of mechanical oscillations, however,
other modes, like w-, g- and r-modes, are not considered here
because their frequencies lie on a very different range and cannot
be detected by the Brazilian and the Dutch detectors, considered in
this work.

\section{Schenberg Bandwidth}

One of the main purposes of this work is to establish the main
properties of the sources when the Schenberg's bandwidth (3.0-3.4
kHz) is considered. From the above discussion the natural candidates
for Schenberg are BHs and NSs.

In the present study, we are particulary concerned with the
determination of the equation of state of NSs. A relevant question
is how to know, in a putative detection by Schenberg, if the source
is  a NS or a BH. The basic way to distinguish them can be through
the damping times of their oscillating modes. The damping time for
the quasi-normal modes of BHs is orders of magnitude shorter than
the f- and p- modes of NSs.

There is a simple relation correlating a Schwarzschild BH mass to
the frequency of its fundamental (quadrupole) quasi-normal mode
(see, e.g., Ref. \cite{kokkotas})

\begin{equation}
\omega=12.07\, {\rm kHz}\, \left(\frac{M_\odot}{M}\right)
\end{equation}
where M is the mass in $M_\odot$. This relation implies that the
Schenberg antenna will only see BHs if their masses are about
$3.5-4.0M_\odot$. It is worth mentioning that if rotation is
considered the above mass range could be extended. A BH of
$9M_\odot$ at maximum rotating velocity could in principle be seen
by Schenberg \cite{costa}.

The damping time of this mode is given by \cite{kokkotas}

\begin{equation}
\tau=0.05\,{\rm ms} \, \left(\frac{M}{M_\odot}\right)\, ;
\end{equation}
which for BHs with masses $3.5\le M\le4.0M_\odot$ gives $\simeq 0.2$
ms.

The same procedure is now applied to identify f- or p-modes in the
Schenberg antenna. A NS f-mode with frequency about 3.0 kHz presents
a damping time much larger then those presented by a BH, being of
the order of 100 ms. This is also the case for the first p-mode,
which would have damping times greater then a few seconds. So, the
differences of a BH ringdown and a NS f- and p-modes are easily
identified by their corresponding damping times.

Even though Schenberg could not determine the damping times within
low errors, such a great difference between these three modes could
enable us to determine the mode that would excite the sphere.

As a restricted bandwidth implied in a corresponding restriction of
BH candidates due to their masses, it shall also imply in a
restriction of NSs candidates. In order to do a model-independent
analysis, we have applied the Schenberg bandwidth into the empirical
relations obtained by Benhar {\it etal} \cite{benhar}. These relations
determine the wave frequency and damping time when the stellar mass
and radius is known.

\begin{figure}
\begin{center}
\includegraphics[width=9cm, angle=0]{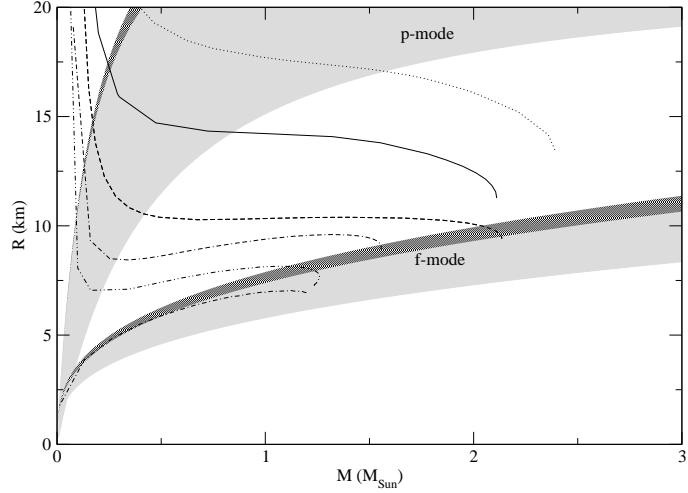}
\caption{\label{1} Mass-radius relation for the Benhar {\it etal}
empirical relation and for different EOSs. The light shaded regions
represent the empirical relations for the f- (lower) and first
p-mode (upper) for the Schenberg bandwidth while the dark ones
describe the additional contribution due to the Mini-GRAIL
bandwidth. The results are compared, for the NL model, to the normal
static (solid line) and (maximum) rotating NS (dotted line) and the
Taurines model with K=220MeV (dashed line). Static strange quark
stars for the MIT bag model are plotted with $B=60MeV/fm^3$
(dot-double dashed line) and $B=100MeV/fm^3$ (dash-double dotted
line) and Chromo-dieletric model (dot-dashed line).}
\end{center}
\end{figure}

We have followed in the opposite direction and determined the NS
masses and radii that would emit f- or p-mode waves in the detector
Schenberg bandwidth. The main results of this analysis are
synthesized in figure \ref{1}, where we have plotted the regions
that correspond to NSs whose GW lies in Schenberg frequency band.
The NS mass-radius relation determines the candidates for a
detection by the Schenberg antenna.

Considering, in equation 1, a bandwidth of 3.0-3.4 kHz we obtain the
mass-radius relation represented by the lower shaded region in
figure 1:

\begin{equation}
R=\left(\frac{33(\pm2)}{3.2(\pm0.2)-0.79(\pm0.09)}\right)^{2/3}M^{1/3} \, \, .
\label{fr2}
\end{equation}

The upper shaded region is obtained through the same procedure,
using the empirical formula for the first pressure mode (equation
2):

\begin{equation}
R=\frac{79(\pm4)M}{3.2(\pm0.2)M+1.5(\pm0.8)} \, \, ,
\end{equation}
where, in either of above equations, the mass  and radius are given
in $km$.

We also plotted in figure 1 a couple of equations of state such as
the Taurines \cite{taurines} and non-linear (NL)\cite{marranghello}
models for nuclear matter, the MIT bag model \cite{mit} and the
Chromo-Dieletric model \cite{malheiro} for quark matter, which we
consider in detail in the next section.

If a f-mode is identified in the detector by its damping time, the
most probable source, considering the results of figure 1, would
correspond to a very compact star with radius smaller than 10 km.
The models that fulfill this condition are models of strange quark
stars. NSs with higher masses can hardly account such a small
radius. However, if such a detection is done, it would lead to
important constraints on the NS matter equation of state, since only
a few models and free parameters sets can account for NSs with
$M\approx 2M_\odot$ and $R\approx 10km$ at the same time. On the
other hand, pressure modes would only be expected to come from
low-mass NSs with very large radii.

From the previous analysis, and considering the fact that low-mass
NSs do not appear in a great number in our galaxy, strange stars are
the most probable sources of f-mode waves to be detected by the
Schenberg antenna. Most of the modern NS models cannot account for
the small radius values imposed by the 3.0-3.4 kHz bandwidth. One
can also consider the fact that RX J1856-3754 \cite{drake} can, in
fact, be a strange quark star, and its inferred values of mass and
radius would imply in a f-mode eigenfrequency about that considered
by the Schenberg bandwidth.

\begin{figure}
\begin{center}
\includegraphics[width=7cm, angle=-90]{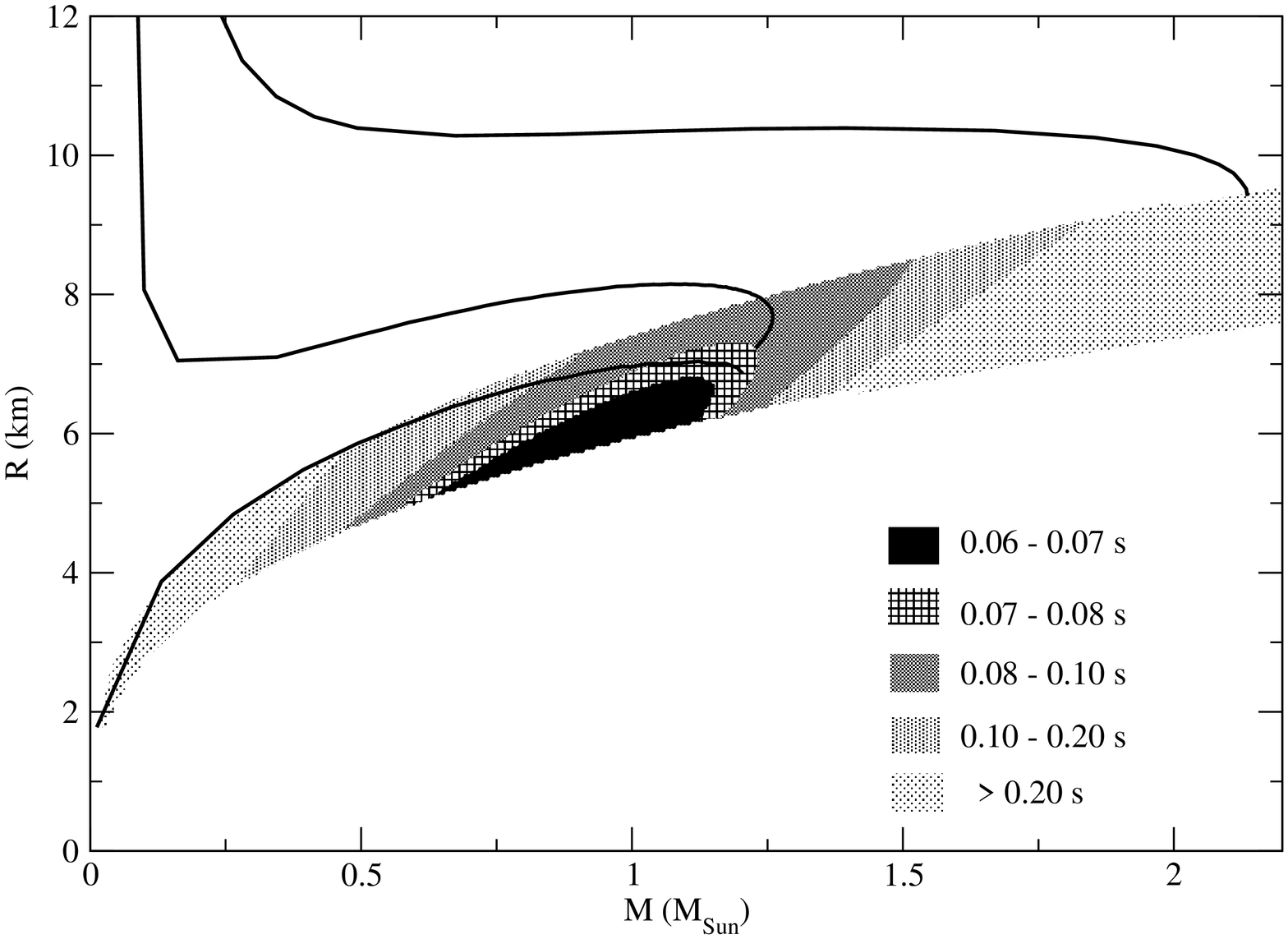}
\vspace{0.5cm} \caption{\label{2} Mass-radius relation for the
Benhar {\it etal} empirical relation. The shaded regions represent the
empirical relations for the f-mode with different damping times.}
\end{center}
\end{figure}

It is worth noting that the constraint given by the Mini-GRAIL
detector is lighter (see \ref{2}), since some other EOSs, of those
considered in the present paper, would contribute to its bandwidth.

Even though the Schenberg spherical antenna cannot determine the
properties of damping time with low errors, we applied the empirical
relations obtained by Benhar {\it etal}, as we have done before, to
describe its properties, considering the NSs in which the f-mode
frequency lies between 3.0-3.4 kHz. In Ref.\cite{benhar} the authors
found an empirical relation for the f-mode damping time described by

\begin{equation}
\tau_f=\frac{R^4}{cM^3}\left[(8.7\pm0.2)\cdot 10^{-2}+(-0.271\pm0.009)
  \frac{M}{R}\right]^{-1}
\end{equation}
where $R$ is the radius and $M$ is the gravitational mass, both in
units of $km$ and $c$ is the speed of light, $3\cdot 10^5km/s$. In
addition to the frequency equation, Eq.\ref{fr}, that gives rise to
the detectable region in the mass-radius diagram, Eq.\ref{fr2}, we
obtained the diagram drawn in Fig.\ref{2}. This diagram shows that,
in the Schenberg detectable region, the signal obtained from NSs
with very high masses, $M>1.8M_\odot$, would have a damping time of,
at least, $0.2 s$, while very low mass NSs, $M<0.5M_\odot$ would
have damping times $\tau>0.1s$. NSs with masses around $1M_\odot$
would have a wide range of possible damping times, $0.06-0.10 s$.

At this point the relevant questions are: what about if a detection
is done? What kind of information we can get from it? These are the
main questions of our study. We have already seen how we could
determine some of the source properties, such as the mass and
radius, and now we need to consider some discussion concerning the
nuclear matter equation of state to find out what kind of
information all of this together could give. We deal with these
issues in the next section.

\section{Nuclear Matter Equation of State}

We have seen that a possible GW detection by the Schenberg antenna
would impose severe constraints in the properties of its sources,
e.g., the masses and radii of the NSs. Once we have considered these
constraints, we can invert the problem and verify its relevance in
the nuclear matter equation of state parameters. Some attempts have
already been done in this direction \cite{ben} but one of the
purposes of this work is to study the information we can get from
the Brazilian detector.

\begin{figure}
\begin{center}
\includegraphics[width=10cm, angle=0]{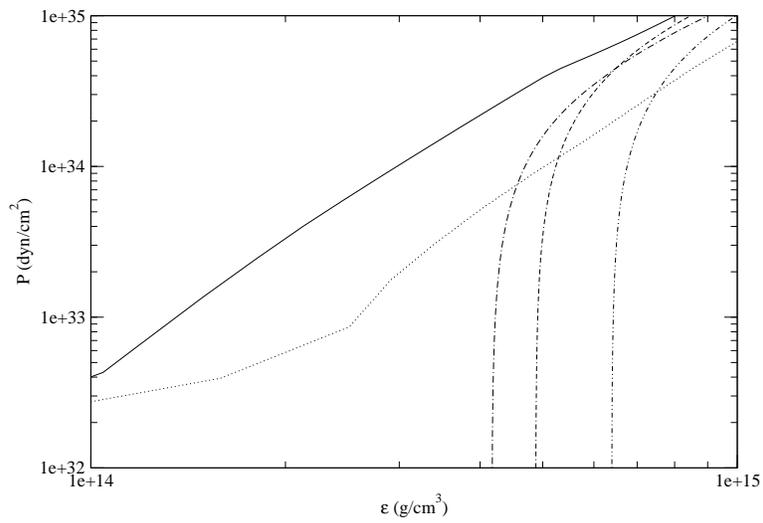}
\caption{\label{3} Equations of State for normal nuclear matter,
using the Taurines model (solid line) and NL model (dotted line),
and for quark matter using the Chromo-dieletric model (dot-dashed
line) and MIT bag model with $B=60 MeV/fm^3$ (dash-double dotted
line) and with $B=100 MeV/fm^3$ (dot-double dashed line). }
\end{center}
\end{figure}

The global properties of mass and radius of compact stars are
directly related to its equation of state. In the previous section
we have developed our analysis in a model-independent level. Here,
we go a step further and, using a specific model for NS matter, we
analyze the consequences of a detection. In particular, we consider
the model developed by Taurines {\it etal} \cite{taurines} for a new
class of parameterized field-theoretical model (see figure \ref{3})
described by the following lagrangian density

\begin{eqnarray}
{\cal L}    &=& \sum\limits_{B}   \bar{\psi}_{B}\left( i\gamma_\mu
(\partial^\mu- g^\star_{\omega B} \omega^{\mu}) -
(M_B-g^\star_{\sigma B} \sigma) - [\frac12 g^\star_{\varrho B}
\mbox{\boldmath$\tau$} \cdot \mbox{\boldmath$\varrho$}^\mu]
\right)\psi_B   \nonumber \\ && +\frac12(\partial_\mu \sigma
\partial^\mu \sigma   - {m_\sigma^2} \sigma^2)  - \frac14
\omega_{\mu \nu}  \omega^{\mu \nu}   + \frac12 {m_\omega^2}
 \omega_\mu \omega^\mu
\nonumber \\ && -   \frac14 \mbox{\boldmath$\varrho$}_{\mu \nu}
\cdot \mbox{\boldmath$\varrho$}^{\mu \nu} +  \frac12m_\varrho^2
\mbox{\boldmath$\varrho$}_\mu  \cdot  \mbox{\boldmath$\varrho$}^\mu
+\sum\limits_{l}   \bar{\psi}_{l} [i \gamma_\mu \partial^\mu   -
M_l] \psi_l  \,\, , \label{eqdl}
\end{eqnarray}
where the baryon field $\psi_B$ is summed over the whole baryon
octet and coupled to the scalar and vector fields $\sigma$, $\omega$
and $\varrho$ through the parametrized coupling constants
$g^\star_{\sigma B}$, $g^\star_{\omega B}$, $g^\star_{\varrho B}$.
The free lepton fields $\psi_\lambda$ contributes to the eletrical
equilibrium in the NS matter. The masses of the baryons, mesons and
leptons are represented by $M_B$, $m_{\sigma,\omega,\varrho}$ and
$m_\lambda$, respectively.

Using a parameterization, $\lambda$, for the baryon-meson coupling
constants,

\begin{equation}
g^\star_\sigma{\bar\psi}\sigma\psi=\frac{g_\sigma\sigma}{\left(1+
    \frac{g_\sigma\sigma}{\lambda M}\right)^\lambda}{\bar\psi}\psi
\end{equation}
\noindent this model describes a wide range of NS parameters, such
as a maximum mass ranging from (very low values) $M=0.66M_\odot$ up
to $M=2.77M_\odot$, and the corresponding radii that vary in the
range of $8<R<13.17\, km$. Each pair in the mass-radius relation is
associated with a different parametrization of the equation of state
(see Ref.\cite{taurines} for further details). Of course, each value
of such parametrization represents different values of nuclear
matter properties.

\begin{figure}
\begin{center}
\includegraphics[width=9cm, angle=-90]{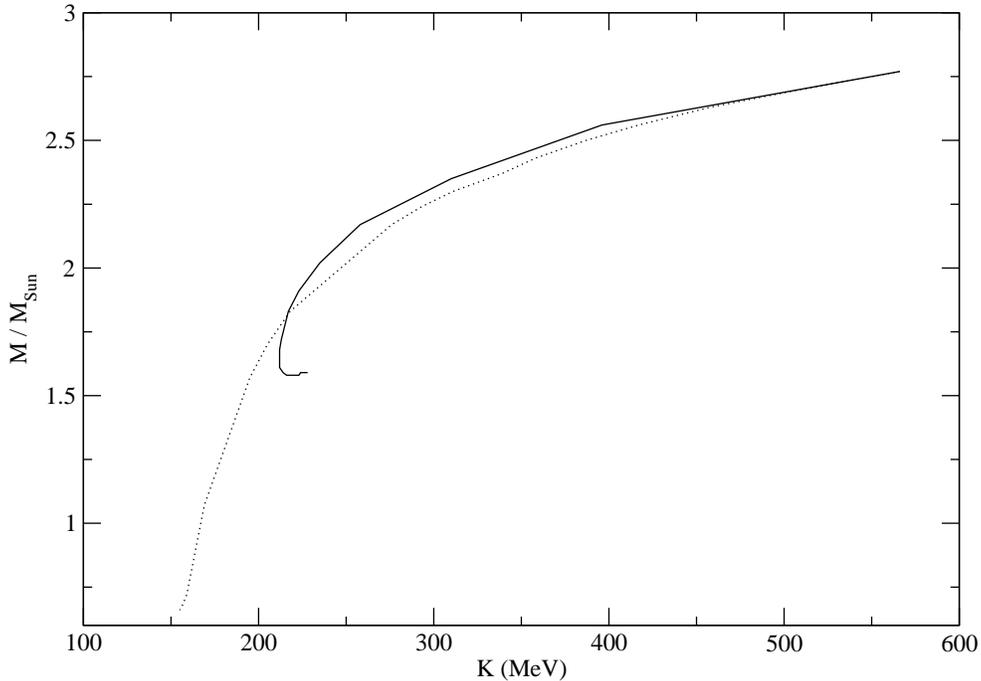}
\caption{\label{4} The NS maximum masses versus the compression
modulus, for different parametrization in the Taurines {\it etal} models.
The solid line represents a parametrization in the scalar mesons
while the dotted line describes a parametrization in both scalar and
vector contributions of the nuclear force.}
\end{center}
\end{figure}

The study of nuclear matter properties is still a very open problem.
We focuse here on the problem of the determination of the
compression modulus, $K$. The compression modulus defines the
curvature of the equation of state or, in a few words, is related to
the capability of matter to be compressed. One can see in figure
\ref{4} that a maximum mass of NSs is obtained for large values of
the compression modulus. The definition of $K$ reads

\begin{equation}
K=9\left[\rho^2\frac{d^2(\epsilon/\rho)}{d\rho^2}\right]_{\rho=\rho_0}
\end{equation}
where $\epsilon$ is the energy density, $\rho$ is the baryon density
and $\rho_0$ the baryon saturation density. Actually, the acceptable
values for the compression modulus are constrained between
$200<K<300MeV$ \cite{blaizot,myers}.

An important constraint to the equations of state is to describe a
family of NSs with a maximum mass greater than the value obtained by
the most massive known pulsar. The PSR J0751+1807 has a mass $M=
2.1\pm 0.2M_\odot$ \cite{nice}. This condition has to be fulfilled
by the set of parameters chosen to describe the NS model. This first
condition already constrains the value of compression modulus to
$K\ge 220 MeV$ when this model is considered. Smaller values of K
would lead to masses smaller than $1.9M_\odot$ disagreeing with the
measures of PSR J0751+1807 mass.

Assuming the detection of a f-mode, associated with a NS composed of
nuclear matter, which we have seen as very unexpected, this would
imply a very compact NS, with high masses ($>1.8M_\odot$) occupying
a small radius, namely, 9-10 km. As the maximum mass star has  the
smallest radius (see figure 1), we look for a set of parameters
which leads to the smallest radius for the maximum mass star.
Stellar models on which the maximum mass has a radius $R\le 10$ km
have compression modulus $215<K< 225 MeV$. Stellar models with
masses smaller then $2.5M_\odot$ and radii greater then 10 km cannot
be detected by the Schenberg antenna, so, a putative detection would
be responsible for the most important constraint on the values of
$K$ even done by any earth-based nuclear experiment, namely,
$220<K<225MeV$.

It was already proposed the existence of low-mass NSs in LMXBs and
HMXBs (see, e.g., Refs. \cite{jonker,orosz}). The detection and
identification of a p-mode wave would confirm this existence. It is
worth mentioning that, as a spherical antenna can be used to
identify the direction of the source, the detection could also
confirm that low-mass NSs are, in fact, related to the LMXBs and
HMXBs.

Another scenario to be discussed concerns the detection of an f-mode
when the source is a strange quark star.
The models used in this analysis are the MIT bag model \cite{mit}
and the chromo-dielectric model \cite{malheiro}. The detection of an
f-mode wave coming from a strange star would impose new constraints
on the values of the bag constant B, the strange quark mass and the
perturbative constant $\alpha_c$.

The MIT bag model describes, as its own name suggests, that baryons
are composed by three quarks confined inside a bag. The effects of
pressure difference in the interior and exterior regions of the bag
are summarized in the bag constant. The thermodynamical potential of
quarks in the MIT bag model is described by

\begin{eqnarray}
\Omega & = & \sum_{q=u,d,s}\frac{-1}{4\pi^2}\left[\mu_q {k_F}_q(\mu_q^2-\frac52
m_q^2)+\frac32 m_q^4 ln\left(\frac{\mu_q+{k_F}_q}{m_q}\right)\right] \nonumber
\\ &+& \frac{2\alpha_c}{4\pi^3}[3\left[\mu_q {k_F}_q-m_q^2-m_q^2
ln\left(\frac{\mu_q+{k_F}_q}{m_q}\right)\right]^2 \nonumber \\
&-&2{k_F}_q^4 - 3m_q^4 ln^2\left(\frac{m_q}{\mu_q}\right)+6
ln\left(\frac{\rho_r}{\mu_q}\right)[\mu_q {k_F}_q m_q^2\nonumber \\
&-&m_q^4
ln\left(\frac{\mu_q+{k_F}_q}{m_q}\right)]]+B
\end{eqnarray}
where $q=u,d,s$ are the three quark flavors up, down and strange, $\mu_q$ is
the chemical potential, $m_q$ represents the quark masses, $k_q$ is the Fermi
momentum of the particles, $B$ is the bag constant and the terms
multiplied by $\alpha_c$ represents the contribution of the first order
perturbation in the strong interaction.

Even with considerable advances in the study of QCD (quantum
chromo-dynamics) on the lattice, from where we expect the most
reliable theoretical results and the consequent advances in
collision experiments, the bag constant still presents a very wide
range of possible values. In the same way, the perturbative
constant, $\alpha_c$, the one that represents the first order
correction to the strong interaction forces between quarks, also
represents an open issue. Astrophysical and GW experiments may
provide a short-cut to the determination of these values.

We argue that, if a GW detection is done, and identified as an f-mode
wave coming from a strange quark star, the higher are the values of
the bag constant, $B$, the perturbative constant, $\alpha_c$ and the
strange quark mass, $m_s$, the more compact is the star and more the
model fits into the region of Schenberg's sensibility curve.

The set of constants, $B=60MeV/fm^3$, $\alpha_c =0$ and $m_s=150MeV$
could not describe stars that emit 3.0-3.4 kHz waves, on the other
hand, $B=100MeV/fm^3$, $\alpha_c=0.35$ and $m_s=200MeV$ would
perfectly fit the optimum region. Our results are in perfect
agreement with those found in Ref.\cite{ben}, where the higher is
the value of B, the higher is the f-mode frequency.

The same behavior appears in the color-dieletric model when we
convert the parameters to obtain the corresponding values of a bag
constant. These constraints in the values of the bag and
perturbative constants would severely restrict the QCD parameters.
If a detection of a strange star is done by the Schenberg antenna
and, consequently, these values are confirmed, the hadron-quark
phase transition would occur in an extremely high density. Such high
density would rule out the possibility of finding standard NSs with
quark cores.

Last, but not least, we refer the reader to Fig. \ref{2}, from
which we can conclude that the determination of the damping time
could also contribute to constraint the parameters of the equations
of state and also the star mass range.

\section{Final remarks}

We have briefly reviewed the properties of GW emission by NSs and
BHs. We also have seen the main characteristics of the Brazilian and
Dutch spherical antennas.

Assuming a possible detection of GWs coming from a NS in the range
of 2.8-3.4 kHz, the one by Schenberg and Mini-GRAIL, we can extract
information to describe the source characteristics. The
identification of the source could be done by means of the wave
damping time, which combined with the corresponding frequency, leads
to the determination of the mass and radius of the star.

The determination of these values imposes important constraints on
the nuclear and subnuclear matter equation of state, such as
limiting values for the compression modulus, nucleon effective mass,
bag and perturbative constants. As a matter of example, values like
$K\approx220MeV$,  $B\approx100MeV/fm^3$ and $\alpha_c\approx0.35$
would be imposed by a putative detection.

It is still important to mention that spherical antennas can
identify the position of the source in the sky, with $20^o$ for the
first stage of the Schenberg antenna, operating at $T\sim4.2K$ and
about $2^o$ at mili-Kelvin temperatures \cite{costa}. This property
would enable us to combine GW data with those obtained by
observations in the electromagnetic spectra, which together could
help identifying the sources.

Last but not least, a relevant question has to do with the event
rate of detectable oscillating modes of NSs within the Schenberg
bandwidth. We have seen in the introduction of the present paper
that there exists a host of possible mechanisms to excite the
oscillating modes of NSs. Many of these mechanisms, however, present
very uncertain event rates. If the f-mode is excited, for example,
during the formation of the NSs, the event rate would be the one
related to the rate of supernovae, which for our Galaxy amounts one
event every tens of years.

In recent studies de Araujo {\it etal} \cite{araujo2005,araujo2006}
consider the event rates associated with pulsar glitches and  also
in LMXBs. They argue that up to several event years would be
possible. We refer the reader to these studies for further details.

\section{Aknoledgments}
GFM and JCNA would like to thank CNPq (grants 381682/2006-4,
303868/2004-0, respectively) for financial support. We also would
like to thank Flavio D'Amico for fruitful and helpful discussions
concerning the X-ray binaries. Last, but not least, we would like to
thank the referees whose suggestions and criticisms greatly improve
our paper.


\end{document}